# Numerical Study of Jet-Like Outwash from Multi-Rotor eVTOLs and Strategies for Flow Mitigation

**Author:** Yen-Chen Chen[1,2,3*], Chih-Che Chueh[1**]

[1]Department of Aeronautics and Astronautics, National Cheng Kung University, Tainan 701, Taiwan
[2]Peniel Infrastructure LTD., 254 Chapman Road, Suite 208, 23642, Newark, Delaware 19702, USA
[3]Department of Physics, Worcester Polytechnic Institute, Worcester, Massachusetts 01609, USA
Corresponding: [*]info@penielinfrastructure.com, [**]chuehcc@mail.ncku.edu.tw

## Abstract

This study presents a comprehensive computational investigation of outwash phenomena generated by electric vertical takeoff and landing (eVTOL) aircraft, with particular emphasis on how rotor geometry and alignment shape hazardous airflow patterns at vertiports. Using nondimensional Reynolds-averaged Navier–Stokes (RANS) simulations with the k–ω SST turbulence model implemented in OpenFOAM, the research systematically characterizes jet-like outwash across a range of multi-rotor configurations. Results demonstrate that propeller count and inter-propeller spacing are primary determinants of outwash intensity, orientation, propagation range, and boundary-layer structure. For power-lift eVTOLs, low propeller counts combined with narrow spacing produce highly directional, intensified jet-like outwash with propagation ranges far exceeding current FAA EB105a safety standards. Conversely, higher propeller counts and larger spacings reduce peak velocities and propagation distances, enabling safer and more compact vertiport layouts. High-propeller-count designs further exhibit vertically stratified and thickened outwash boundary layers, requiring tailored mitigation strategies. Targeted engineering solutions—such as modular blast deflectors aligned with predicted outwash directions—are shown to reduce required vertiport safety areas by up to 82% without compromising operational safety. These findings establish a direct link between aircraft design, regulatory compliance, and infrastructure optimization, offering practical pathways for safe and scalable urban air mobility. The study also provides a foundation for future research on optimal mitigation device geometries and system-level integration of eVTOL operations within urban environments.

# 1. Introduction

The rapid development of electric vertical takeoff and landing (eVTOL) systems heralds a transformative era in urban air mobility and regional transportation. However, this technological leap also introduces critical challenges in managing aerodynamic interactions with ground infrastructure, particularly at vertiports. Among these, outwash—the strong outward jet-like airflow generated by distributed lift rotors during takeoff and landing—has emerged as a dominant concern for both safety and sustainability [1–6]. Uncontrolled outwash poses risks that extend beyond vehicle performance: it can generate ground-level wind velocities exceeding pedestrian safety limits, disrupt and damage surrounding equipment, and demand excessively large safety buffers around vertiports. Such requirements could undermine land-use efficiency, compromise economic feasibility, and erode public acceptance of urban air mobility [7–12]. Consequently, effective outwash mitigation is not merely a design optimization, but a prerequisite for ensuring the safe, practical, and socially sustainable integration of eVTOL systems into modern transportation networks.

Regulatory agencies, including the FAA and the UK Civil Aviation Authority (CAA), increasingly acknowledge that outwash generated by eVTOL aircraft is substantially more intense, anisotropic, and spatially extensive than that of conventional single-rotor helicopters [13–16]. Both field experiments and high-fidelity CFD simulations demonstrate that the geometry, number, and arrangement of eVTOL propellers critically determine the strength, directionality, and spatial extent of outwash flows—effects that are especially pronounced in densely packed multi-rotor architectures envisioned for advanced air mobility systems [17–19]. In response, updated regulatory frameworks such as FAA Engineering Brief No. 105A [16] (the "EB105a") now mandate the designation of downwash/outwash caution areas (DCAs) for vertiports whenever ground-level wind velocities exceed prescribed safety limits (34.5 mph). Without effective mitigation measures, these requirements could drastically expand vertiport land footprints, threatening the practicality and scalability of urban air mobility unless novel outwash suppression strategies or infrastructure adaptations are implemented.

Recent empirical evidence—including the 2023 UKCAA project [20] and the FAA's ongoing eVTOL Downwash and Outwash Surveys [3]—has shown that traditional helicopter-based models substantially underestimate the extent of eVTOL outwash hazards. Measurements reveal that hazardous outwash can propagate more than 120 ft beyond the rotor disk and often manifests as impulsive jet-like "vortex bullets," posing acute risks to ground operations and bystanders. Advanced computational studies have further illuminated these phenomena. For example, Brown [1] demonstrated that aligned propeller streams can merge to form high-momentum fountain flows and hazardous lateral gusts, with spatiotemporal structures strongly dependent on aircraft design and ground proximity. Calderone et al. [6], through a systematic review of unmanned aerial spraying systems, highlighted that multi-rotor UAVs generate concentrated outwash streams whose patterns vary with propulsion geometry and operating conditions. Similarly, Stokkermans et al. [14] identified that aerodynamic interactions between rotors in typical eVTOL layouts significantly modify thrust, power, and force characteristics, with side-by-side and tandem configurations both introducing performance penalties and complex slipstream effects sensitive to rotor spacing. Kim et al. [10]

advanced conceptual eVTOL designs by optimizing rotor placement in distributed propulsion, showing through CFD that staggered rotor layouts reduce some wake instabilities yet still generate outwash beyond regulatory assumptions. High-fidelity multi-rotor simulations by Ventura and Yoon [21], employing actuator disc and line methods, confirmed that vortex interactions and jet mergers evolve dynamically with ground effect, rendering simple exclusion-zone models inadequate.

Outwash mitigation has therefore emerged as a critical priority for ensuring safe, sustainable, and scalable vertiport design. Current research and practice explore a range of technologies, including modular outwash barriers, elevated landing pads, ground-effect flow spoilers, and computationally optimized vertiport layouts that suppress hazardous wind propagation [22,23]. For example, Peniel Infrastructure introduced modular barrier systems as an adaptive strategy for urban vertiports, employing site-specific computational analyses to confine outwash streams and reduce impacts on surrounding infrastructure. Maksoud et al. [23] advanced this approach with a computational optimization framework that refines vertiport placement and geometry, integrating high-fidelity simulations to achieve safe separation of outwash zones while supporting operational scalability. Similarly, Urban-Air Port implemented elevated landing platforms and flow-spoiler devices, demonstrating measurable reductions in transient gust formation during eVTOL operations and recommending design refinements to accommodate long-term scalability. Collectively, these innovations highlight the growing importance of aerodynamic management as a cornerstone of vertiport engineering.

The absence of effective outwash mitigation strategies for eVTOL operations risks perpetuating reliance on generic or outdated design practices, which in turn necessitate oversized safety buffer zones, inefficient land allocation, and prohibitively high vertiport construction costs [3,4,16,20]. Without simulation-driven insights into outwash dynamics and tailored mitigation approaches, planners and manufacturers may overlook critical hazards arising from specific rotor geometries and spacing configurations. Such oversights not only compromise safety and increase the vulnerability of surrounding infrastructure to outwash-induced stresses, but also impose stricter operational constraints and material demands. Ultimately, neglecting these considerations threatens to undermine the scalability of urban air mobility frameworks, constraining both their economic viability and infrastructural feasibility in real-world deployment.

There is a pressing need for rigorous physical and computational investigations that explicitly link propeller alignment to far-field metrics of eVTOL outwash, thereby guiding the development of mitigation devices in line with evolving regulatory frameworks. To address this gap, the present study employs high-fidelity CFD simulations to systematically examine outwash behavior across a range of propeller geometries and alignments. The analysis provides a comprehensive assessment of how rotor configuration governs flow structures, hazard zones, and the spatial extent of jet-like outwash. Building on these insights, we propose targeted mitigation strategies and offer design recommendations for eVTOL propulsion systems that minimize hazardous outwash effects through optimized rotor placement and spacing. Rather than focusing primarily on the root causes of these phenomena, our approach emphasizes the interpretation of observed flow behaviors to directly extract practical design guidance. This strategy ensures that the findings can be translated into actionable engineering solutions for vertiport operations. Ultimately, the results demonstrate that carefully engineered interventions—minimal, cost-effective, and structurally feasible—can substantially reduce outwash buffer-zone requirements, shrinking vertiport footprints and material demands without compromising operational safety.

# 2. Methodology

## 2.1 *Numerical model for eVTOL outwash simulation*

In this study, the simulation of eVTOL outwash is conducted within a nondimensional framework, in which all governing equations are normalized using appropriate characteristic scales. The characteristic length is defined by the propeller diameter $D$ (Fig. 1a), while the characteristic velocity is taken as the rotor disk flow velocity $U$, computed as

$$U = \sqrt{\frac{(T/A)}{2\rho}}, \tag{1}$$

where $T$ denotes the thrust generated by a single propeller, $A$ represents the cross-sectional area of the propeller outlet, and the ratio $T/A$ corresponds to the disk loading, a key parameter that characterizes the aerodynamic performance of the propeller system.

This nondimensional framework enables consistent scaling and meaningful comparison across propellers of varying sizes and operating conditions by expressing the governing dynamics through dimensionless groups. Within this context, the Reynolds-averaged Navier–Stokes (RANS) equations for incompressible flow can be reformulated in nondimensional form (Swanson [24]) as:

$$\frac{\partial \tilde{u}_i}{\partial \tilde{x}_i} = 0, \tag{2}$$

$$\frac{\partial \tilde{u}_i}{\partial \tilde{t}} + \tilde{u}_j \frac{\partial \tilde{u}_i}{\partial \tilde{x}_j} = -\frac{\partial \tilde{p}}{\partial \tilde{x}_i} + \frac{1}{Re_D}\frac{\partial^2 \tilde{u}_i}{\partial \tilde{x}_j^2} - \frac{\partial \tilde{u}_i' \tilde{u}_j'}{\partial \tilde{x}_j}, \tag{3}$$

where all spatial coordinates are normalized by the propeller diameter $D$, velocities by the characteristic velocity $U$, time by $D/U$, and pressure by $\rho U^2$. For clarity, the tilde symbol ˜ is used to denote nondimensionalized quantities, while the overbar ¯ indicates mean values. The corresponding Reynolds number is defined as $Re_D = \rho U D/\mu$ . Turbulence closure is achieved using the nondimensionalized $k$–$\omega$ SST model, with the associated transport equations expressed (Menter [25]) as:

$$\frac{\partial \tilde{k}}{\partial \tilde{t}} + \tilde{u}_j \frac{\partial \tilde{k}}{\partial \tilde{x}_j} = \tilde{P}_k - \tilde{\beta}^* \tilde{k}\tilde{\omega} + \frac{1}{Re_D}\frac{\partial}{\partial \tilde{x}_j}\left[(\tilde{\nu} + \tilde{\sigma}_k \tilde{\nu}_t)\frac{\partial \tilde{k}}{\partial \tilde{x}_j}\right], \tag{4}$$

$$\frac{\partial \tilde{\omega}}{\partial \tilde{t}} + \tilde{u}_j \frac{\partial \tilde{\omega}}{\partial \tilde{x}_j} = \alpha \frac{\tilde{\omega}}{\tilde{k}} \tilde{P}_k - \beta \tilde{\omega}^2 + \frac{1}{Re_D}\frac{\partial}{\partial \tilde{x}_j}\left[(\tilde{\nu} + \tilde{\sigma}_\omega \tilde{\nu}_t)\frac{\partial \tilde{\omega}}{\partial \tilde{x}_j}\right] + 2(1 - F_1)\frac{\tilde{\sigma}_{\omega 2}}{\tilde{\omega}}\frac{\partial \tilde{k}}{\partial \tilde{x}_j}\frac{\partial \tilde{\omega}}{\partial \tilde{x}_j}, \tag{5}$$

By recasting the governing flow and turbulence equations in nondimensional form, the underlying physics of propeller outwash can be examined independently of specific rotor scales. This approach ensures that the findings retain broad generality, enabling meaningful application and comparison across diverse eVTOL configurations and design scenarios.

In this study, a normalized Cartesian coordinate system is adopted for the simulations, where $\tilde{x}$ and $\tilde{y}$ denote the ground-plane directions and $\tilde{z}$ represents the vertical height. The eVTOL is modeled in hover at $\tilde{z} = 1$ within a rectangular computational enclosure. Boundary conditions are defined as follows: the top surface is specified as a pressure inlet, the four side surfaces as pressure outlets, and the bottom surface as a no-slip ground boundary. This setup follows the configuration illustrated in Fig. 3 of Suchanek and Filipek [26]. A turbulence intensity of 20% is imposed to represent the highly turbulent conditions characteristic of eVTOL outwash (Riccobene et al. [27] and Barber et al. [28]). The CFD analysis is performed in OpenFOAM under steady-state conditions, with the actuator disk model employed to capture propeller-induced effects [29]. Each propeller blade is modeled using a NACA 4412 airfoil profile, chosen for its favorable lift characteristics at low-to-moderate angles of attack—operating regimes typical of drone and eVTOL propulsion systems [30,31]. Mesh independence was systematically verified, with all key results and flow features shown to be insensitive to further grid refinement.

To further justify the use of the steady-state $k$–$\omega$ SST model in this study, it is essential to emphasize its distinctive advantages. By combining the robustness of the $k$–$\omega$ formulation in the near-wall region with the free-stream independence of the $k$–$\varepsilon$ model, the $k$–$\omega$ SST approach ensures reliable resolution of boundary-layer flows while offering improved accuracy in predicting flow separation in external aerodynamic applications (Menter [25] and Menter [32]). This hybrid capability has made it a widely adopted standard in engineering practice, particularly for problems characterized by strong pressure gradients, complex boundary layers, and moderate separation—conditions commonly encountered in aircraft and rotorcraft aerodynamics (Younoussi and Ettaouil [33]). In the context of eVTOL operations, the steady-state $k$–$\omega$ SST model provides a practical and industry-recognized framework for assessing the average effects of turbulent outwash on vertiport surfaces and aircraft structures, yielding insights into surface forces, safety buffer zones, and regulatory compliance requirements (Qiao et al. [34]).

## 2.2 *Propeller configurations of eVTOL*

To systematically examine the outwash characteristics associated with different eVTOL propulsion layouts, this study simulates the aerodynamic behavior of seven representative propeller configurations, as illustrated in Figure 1. The analysis spans a range of propeller counts ($P$), including a single-rotor helicopter ($P = 1$; Fig. 1a), a quadcopter ($P = 4$; Fig. 1b), a hexacopter ($P = 6$; Fig. 1c), and several power-lift eVTOL concepts with $P = 2,6,8,$ and 12 propellers (Fig. 1d–g) distributed along the wingspan ($x$-direction). Each configuration is evaluated under standardized propeller spacing conditions, with nondimensional parameters $\tilde{L} = L/D$ and $\tilde{W} = W/D = 3$. For multi-rotor cases, adjacent propellers are arranged to counter-rotate, thereby capturing realistic aerodynamic interactions. The operational boundaries of the Touchdown and Liftoff Area (TLOF) and the Final Approach and Takeoff Area (FATO) for each configuration are also delineated in the figure. Following FAA's EB105a [16], the TLOF diameter is defined by the rotor diameter of the eVTOL ($RD$), corresponding to the smallest circle that fully encompasses all propellers, and is calculated as:

$$\tilde{RD} = RD/D = \begin{cases} 1; P = 1\ (a) \\ \sqrt{2}\tilde{L} + 1; P = 4\ \ (b) \\ 2\tilde{L} + 1\ ; P = 6\ \ (c) \\ \tilde{L} + 1; P = 2\ \ (d) \\ \sqrt{4\tilde{L}^2 + \tilde{W}^2} + 1; P = 6\ \ (e) \\ \sqrt{9\tilde{L}^2 + \tilde{W}^2} + 1; P = 8\ \ (f) \\ \sqrt{25\tilde{L}^2 + \tilde{W}^2} + 1; P = 12\ \ (g) \end{cases}, \tag{6}$$

where the diameter of FATO is defined as twice $RD$, and thus twice the size of TLOF [16].

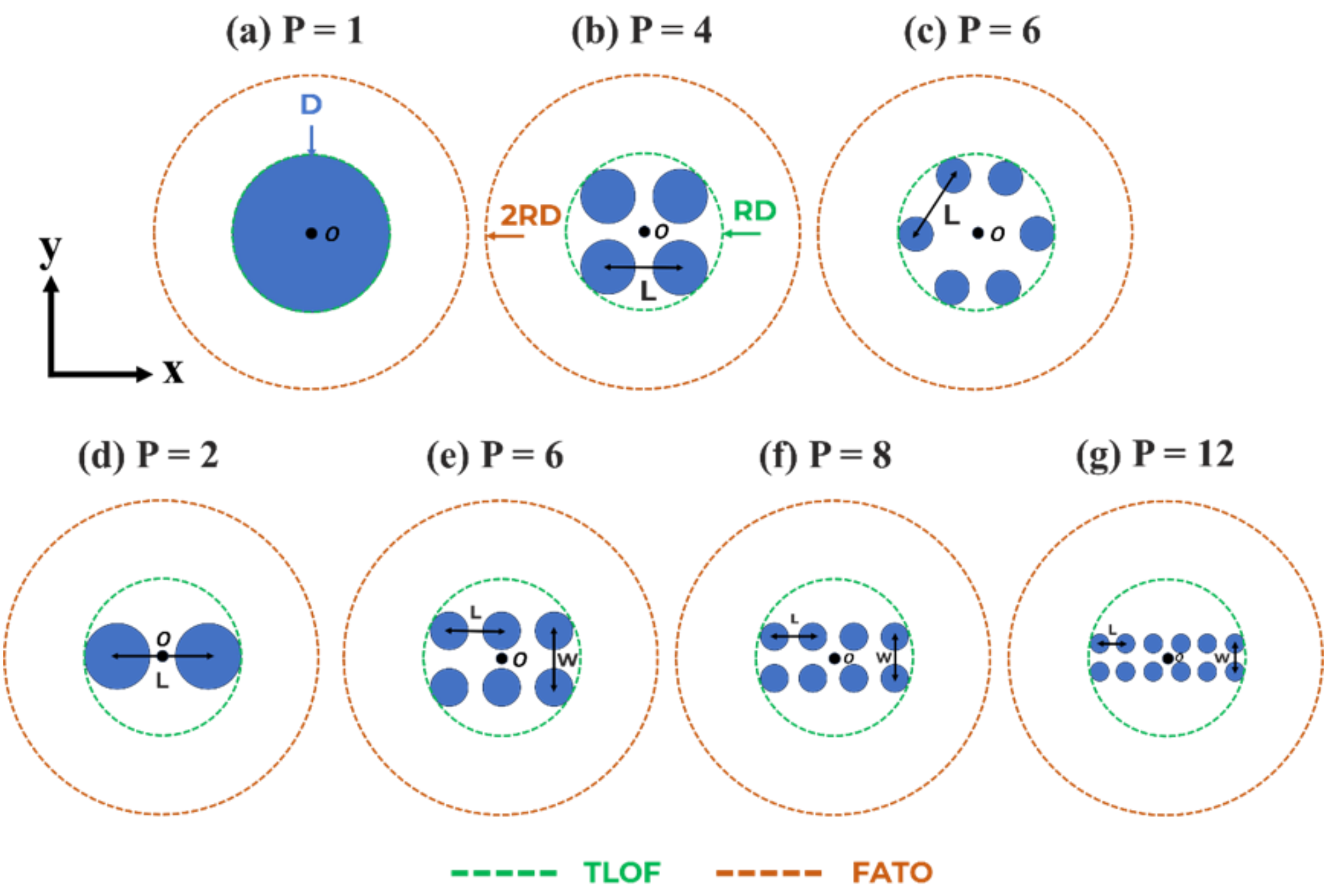

Figure 1 – Geometric configurations of propeller arrangements: (a) helicopter ($P = 1$); (b) quadcopter ($P = 4$); (c) hexacopter ($P = 6$); (d) power-lift eVTOL ($P = 2$); (e) power-lift eVTOL ($P = 6$); (f) power-lift eVTOL ($P = 8$); and (g) power-lift eVTOL ($P = 12$). For the power-lift eVTOL designs, propellers are distributed along the wingspan ($x$-direction). The boundaries of TLOF and FATO are also indicated.

To isolate the effect of configuration geometry on propeller outwash, the total thrust ($T$) is fixed at 30,000 N, while the disk loading is maintained at $T/A = 757$ Pa. As a result, both the propeller diameter ($D$) and rotational speed ($\Omega$) are scaled following the methodology outlined in [20]

$$\frac{D_a}{D_b} = \sqrt{\frac{P_b}{P_a}}, \quad \frac{\Omega_a}{\Omega_b} = \sqrt{\frac{P_a}{P_b}}, \tag{7}$$

# 3. Results and discussion

Section 3.1 presents a comparison between jet-like outwash flows generated by multi-rotor eVTOLs and those of helicopters under matched geometric and operating conditions. Section 3.2 examines the mechanisms of jet formation and the orientational distribution of outwash as shaped by inter-propeller spacing. In Section 3.3, orientation-dependent outwash patterns are analyzed across multiple eVTOL configurations, illustrating how collective jet streams align, merge, and propagate. Section 3.4 quantifies the hazardous propagation range of outwash as a function of nondimensional inter-propeller distance ($\tilde{L}$), highlighting safety risks that extend beyond current vertiport standards. Section 3.5 investigates the boundary-layer structure of outwash at ground level, providing critical insights for the design of mitigation devices. Finally, Section 3.6 evaluates a series of blast deflector concepts, demonstrating that geometry-matched devices can substantially attenuate hazardous outwash, reducing it to FAA EB105a-compliant levels and offering actionable guidance for vertiport safety and scalability.

## *3.1 Multi-rotor eVTOL generates jet-like outwash flow*

Figure 2 presents the distribution of outwash flow speeds for a conventional helicopter ($P = 1$; Fig. 2a) and a power-lift eVTOL configuration ($P = 2$, $\tilde{L} = 1.2$; Fig. 2b), both evaluated at the plane $\tilde{z} = 0.2$, $\tilde{y} = 0$, and $\tilde{x} = 0$. For the helicopter case, the outwash velocity decreases smoothly and uniformly with increasing distance from the rotor in both the $\tilde{x}$ (wingspan) and $\tilde{y}$ (longitudinal) directions (Fig. 2a) [4]. By contrast, the power-lift eVTOL generates a distinctly jet-like outwash originating from the gaps between adjacent propellers (Fig. 2b). A closer comparison of velocity profiles reveals that longitudinal outwash ($\tilde{x} = 0$) is substantially stronger than wingspan-directed outwash ($\tilde{y} = 0$) in the eVTOL case, with the propagation range in the longitudinal direction greatly exceeding that of the conventional helicopter. These findings highlight that multi-rotor eVTOL outwash dynamics are highly sensitive to propeller geometric alignment, particularly the spacing and configuration of inter-propeller gaps. Moreover, the eVTOL outwash exhibits pronounced jet-like behavior, with nondimensional velocities exceeding unity ($\tilde{U} > 1$), indicating that outwash streams can surpass the rotor disk flow velocity and even intensify as they propagate along the ground.

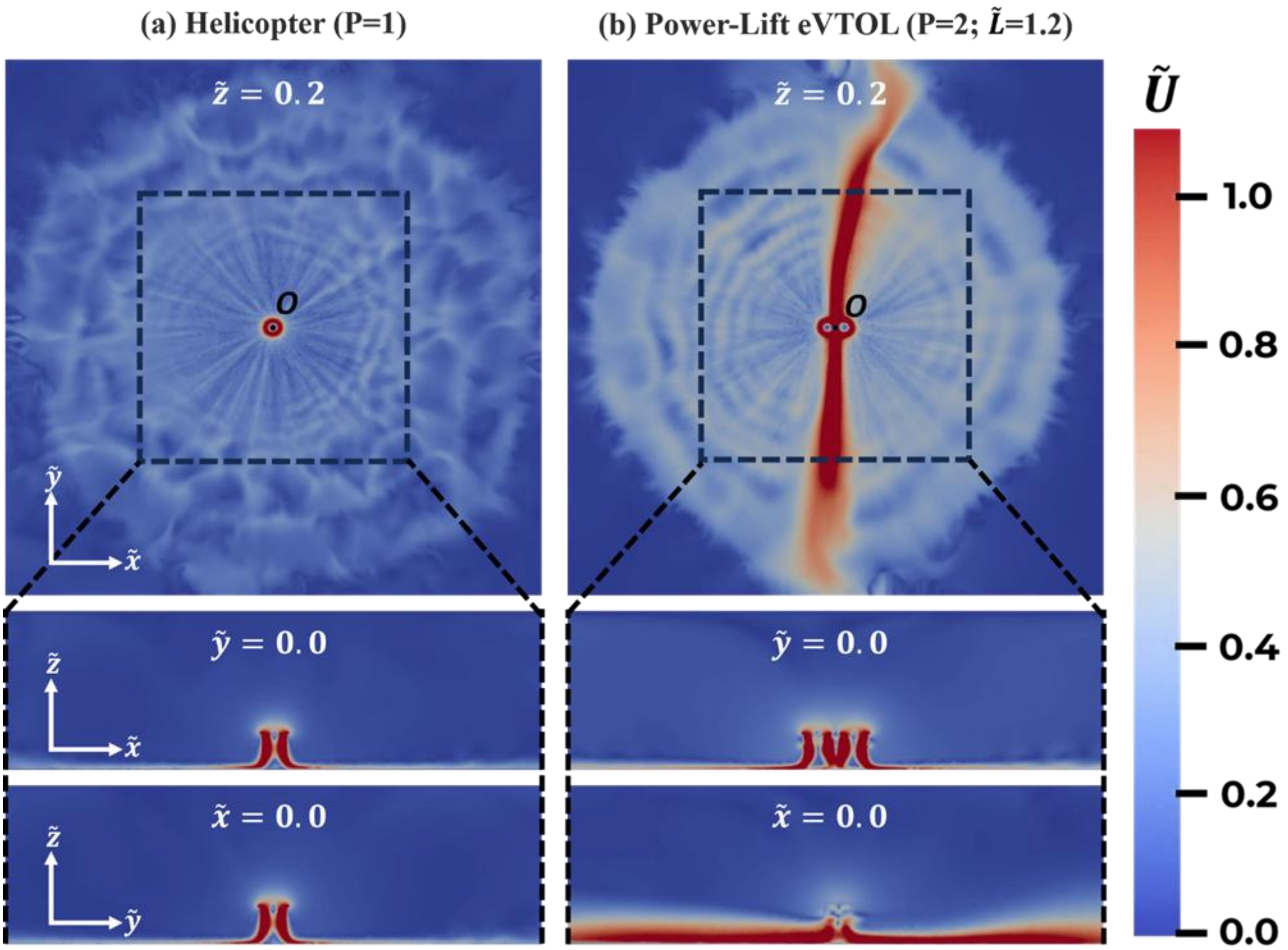

Figure 2 – Outwash flow speed distribution for (a) a conventional helicopter ($P = 1$) and (b) a power-lift eVTOL configuration ($P = 2$, $\tilde{L} = 1.2$), evaluated at the plane $\tilde{z} = 0.2$, $\tilde{y} = 0$, and $\tilde{x} = 0$.

Figure 3 shows the nondimensional pressure distribution ($\tilde{p}$) at the plane $\tilde{z} = 0.2$ (Figs. 3a and 3c) and the corresponding pressure gradients ($\nabla\tilde{p}$) along the $\tilde{x} - \tilde{x}$ and $\tilde{y} - \tilde{y}$ sections (Figs. 3b and 3d) for the helicopter ($P = 1$) and the power-lift eVTOL configuration ($P = 2$, $\tilde{L} = 1.2$), respectively. In the helicopter case, the pressure field decays smoothly with distance from the rotor in both $\tilde{x} - \tilde{x}$ and $\tilde{y} - \tilde{y}$ directions (Fig. 3a), and the pressure gradients along the two sections follow similar trends (Fig. 3b). By contrast, the power-lift eVTOL exhibits a distinct pressure peak at the inter-propeller gap, driven by the convergence of the two downwash jets upon ground interaction (Fig. 3c). Gradient analysis (Fig. 3d) further reveals that the maximum pressure gradient along the $\tilde{y} - \tilde{y}$ section is 1.8 times larger than that along the $\tilde{x} - \tilde{x}$ section. This asymmetry underscores the fundamental mechanism behind the jet-like outwash observed in the longitudinal direction of the multi-rotor eVTOL, as the pressure gradient is the principal driver of fluid motion according to the Navier–Stokes equations.

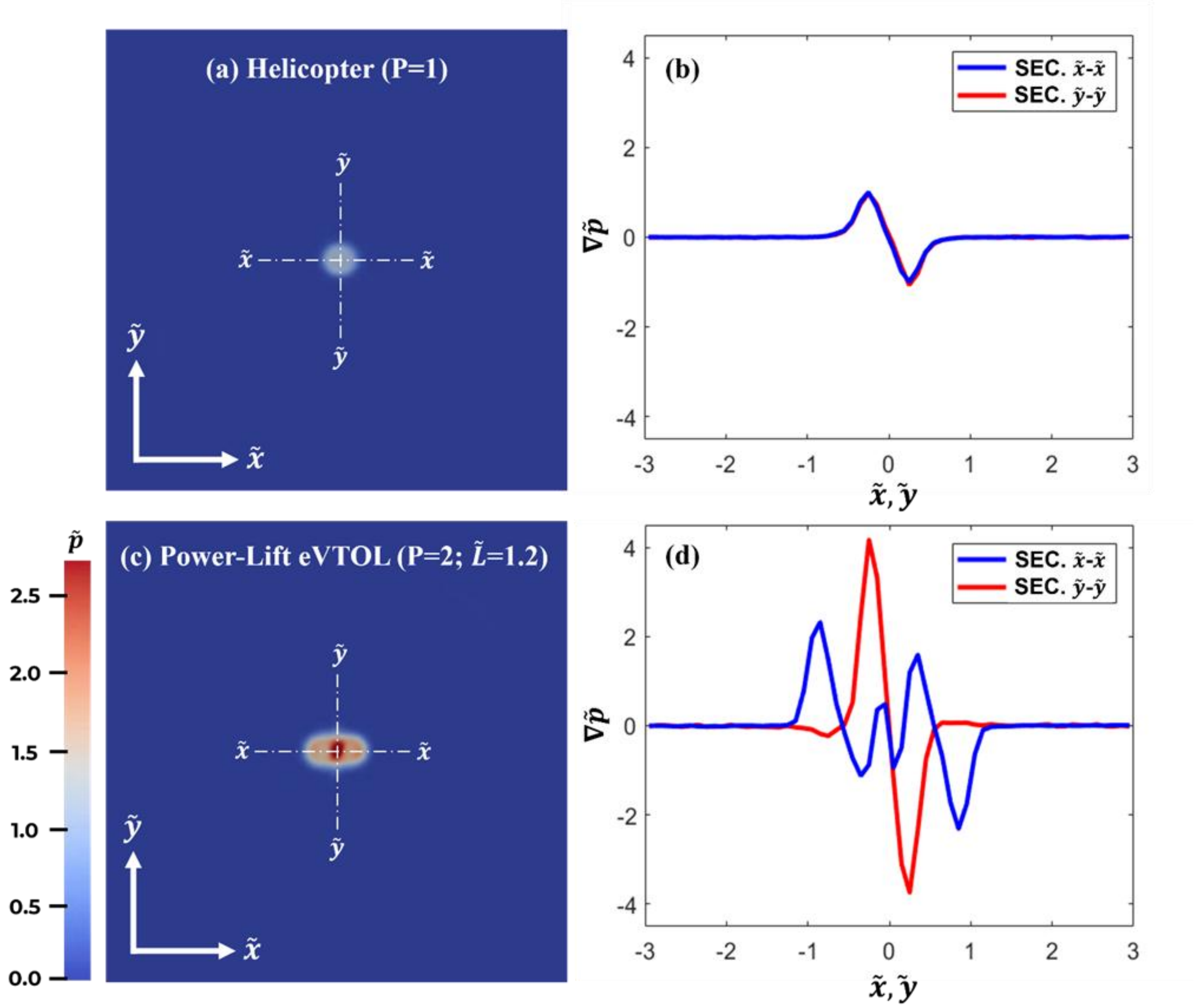

Figure 3 – Pressure fields and gradients for different configurations: (a) pressure distribution of the helicopter ($P = 1$) at $\tilde{z} = 0.2$; (b) corresponding pressure gradients along the $\tilde{x} - \tilde{x}$ and $\tilde{y} - \tilde{y}$ sections from (a); (c) pressure distribution of the power-lift eVTOL ($P = 2$, $\tilde{L} = 1.2$) at $\tilde{z} = 0.2$; and (d) corresponding pressure gradients along the $\tilde{x} - \tilde{x}$ and $\tilde{y} - \tilde{y}$ sections from (c).

The jet-like outwash observed between the propeller gaps of multi-rotor eVTOLs arises from the coupled effects of several aerodynamic phenomena: the mutual interaction of adjacent rotor downwashes, the amplification of ground effect, and the generation and evolution of vortex structures. Each rotor produces its own downwash, and when multiple rotors are placed in close proximity, their flows interact and merge, creating concentrated high-velocity jets in the inter-propeller regions [14,21]. Upon impingement with the ground, these downward flows are redirected laterally; the ground effect intensifies this redirection, confining momentum along the surface and accelerating coherent radial jets between propellers [13,27,28]. In addition, vortex rings shed from rotor tips and lifting surfaces interact with each other and with ground-constrained flows, further energizing the outwash and producing unsteady, spatially complex patterns observed in both experimental and computational studies [21,26,27]. Collectively, these mechanisms—multi-rotor aerodynamic interaction, ground-effect enhancement, and vortex dynamics—explain not only the persistence and intensity of jet-like outwash but also why its propagation exceeds that of conventional single-rotor configurations.

From these insights, several critical research questions emerge:

- What are the dominant factors controlling the formation of jet-like outwash flows?
- How does the spatial orientation of outwash vary with different eVTOL propeller alignment geometries?
- Which near-ground boundary-layer phenomena sustain the long-distance propagation of jet-like outwash?

These questions are addressed systematically in the following sections.

### *3.2 Formation of jet-like eVTOL outwash*

To elucidate the mechanisms driving the formation of jet-like outwash, Figure 4 presents the outwash flow speed distributions for power-lift eVTOL configurations with two propellers ($P = 2$), evaluated across a range of normalized inter-propeller distances ($\tilde{L} = 1.2, 5, 20$; Figs. 4a–c). Complementarily, Figure 4d illustrates the orientational distribution of the mean outwash velocity ($\bar{\tilde{U}}$) within the FATO region, spanning ($\widetilde{RD}/2 \leq \tilde{r} \leq \widetilde{RD}$). The results clearly demonstrate that smaller inter-propeller distances produce markedly stronger jet-like outwash, confirming $\tilde{L}$ as a key geometric parameter governing outwash intensity. Orientational analysis (Fig. 4d) further reveals that the mean outwash velocity aligns predominantly with the directions of the inter-propeller gaps, concentrated at $\tilde{\theta} = 90°$ and $270°$. As the inter-propeller spacing decreases, the magnitude of these jets intensifies; by contrast, at $\tilde{L} = 20$ (Fig. 4c), the jet-like structure effectively vanishes. This finding implies that rotor centers would need to be separated by approximately twenty times the propeller diameter to eliminate jet-like outwash—an unrealistic requirement for practical eVTOL configurations. Collectively, these results highlight that jet-like outwash is not an incidental feature but an inherent characteristic of multi-rotor eVTOL aerodynamics. A rigorous understanding of this mechanism is therefore essential for devising targeted and effective mitigation strategies that ensure safe and efficient vertiport operations.

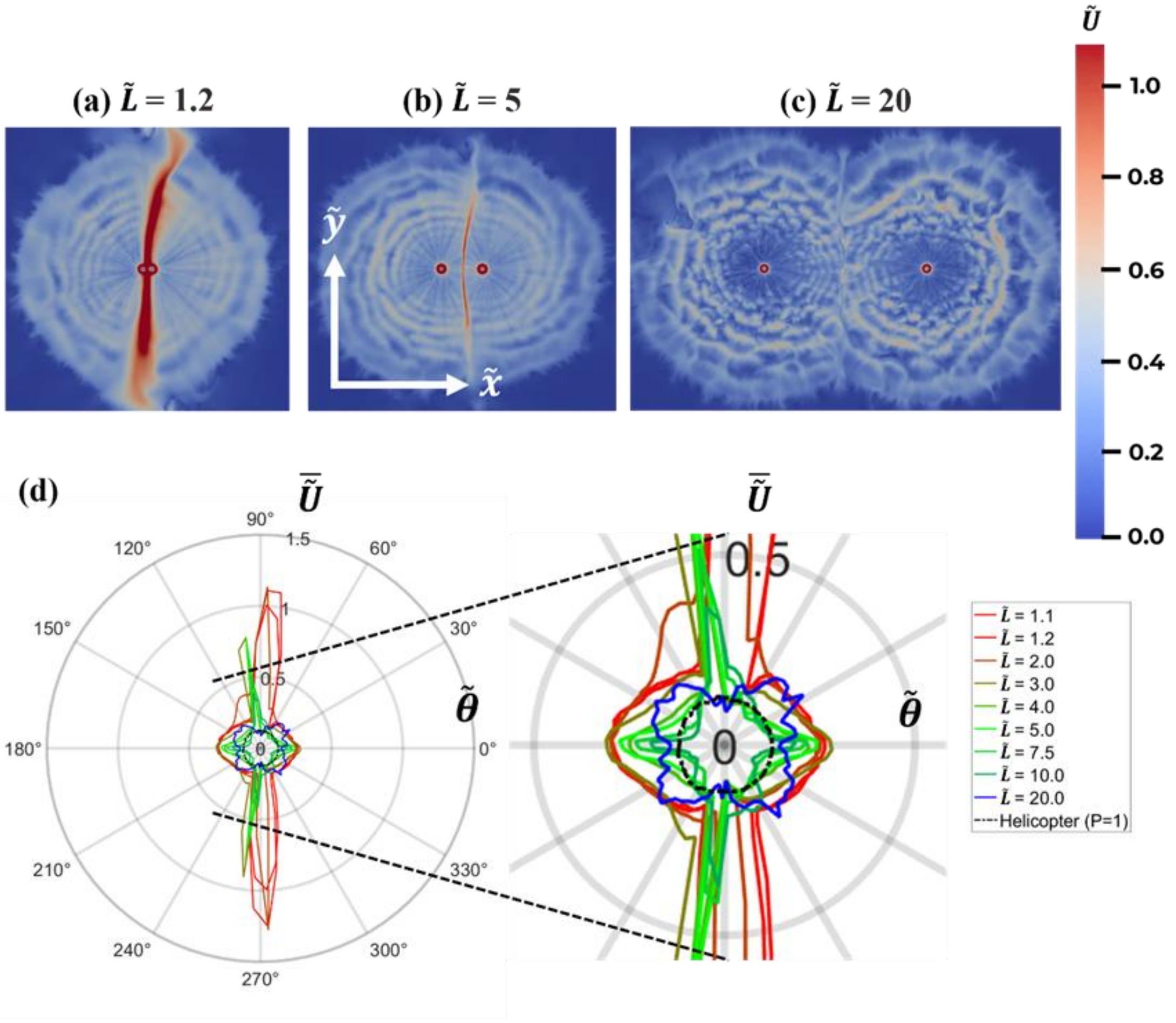

Figure 4 – Outwash flow speed distribution of the power-lift eVTOL ($P = 2$) at a height of $\tilde{z} = 0.2$ for different normalized inter-propeller distances: (a) $\tilde{L} = 1.2$; (b) $\tilde{L} = 5$; (c) $\tilde{L} = 20$. (d) Orientational distribution of mean outwash speed ($\bar{\tilde{U}}$) within the FATO region, spanning $\tilde{L} = 1.1$ to $\tilde{L} = 20$. Here, $\tilde{\theta} = 0°$ corresponds to the $\tilde{x}$-axis direction.

### *3.3 Orientational distribution of jet-like eVTOL outwash*

To deepen the understanding of the orientational characteristics of jet-like outwash across different eVTOL propeller architectures, Figure 5 presents the outwash flow speed distributions for a quadcopter ($P = 4$; Fig. 5a), hexacopter ($P = 6$; Fig. 5b), and power-lift eVTOLs with varying propeller counts ($P = 2,6,8,12$; Figs. 5c–f), all evaluated at a normalized inter-propeller spacing of $\tilde{L} = 1.2$. Complementarily, Figure 6 illustrates the orientational distributions of mean outwash speed ($\bar{\tilde{U}}$) within the FATO region for the same configurations: quadcopter ($P = 4$; Fig. 6a), hexacopter ($P = 6$; Fig. 6b), and power-lift eVTOLs ($P = 2,6,8,12$; Figs. 6c–f). These analyses cover normalized inter-propeller spacings from $\tilde{L} = 1.1$ to $\tilde{L} = 5$, with the exception of the $P = 2$ configuration, which is further extended to $\tilde{L} = 20$ (Fig. 6c).

The results presented in Figures 5 and 6 demonstrate that jet-like outwash flows predominantly emerge along inter-propeller gaps, highlighting the strong dependence of orientational distribution on geometric propeller alignment. For quadcopters and hexacopters, this alignment manifests as four and six distinct jet-like streams, respectively. In contrast, power-lift eVTOL configurations with $P = 6,8,12$ exhibit a markedly different behavior: the outwash flows converge into a collective stream concentrated along the longitudinal axis ($\tilde{\theta} = 90°$ and $270°$) rather than dispersing into multiple independent jets. This convergence results in significantly higher outwash velocities in the longitudinal ($\tilde{y}$) direction compared to the wingspan ($\tilde{x}$) direction. These findings indicate that multiple jet-like flows can aggregate into dominant, orientation-specific streams, offering valuable predictive insight into outwash behavior and informing targeted mitigation strategies for vertiport operations.

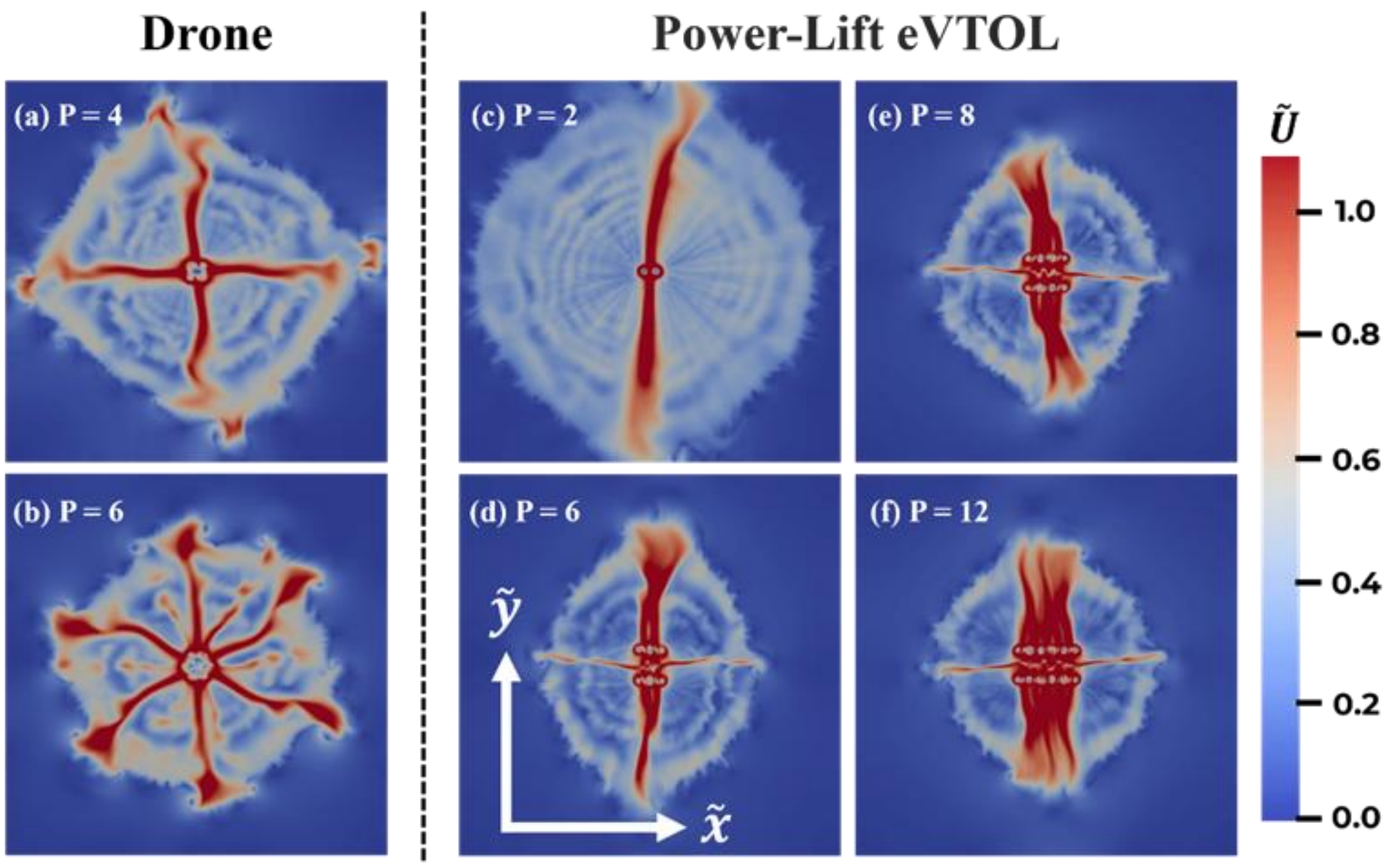

Figure 5 – Outwash flow speed distributions at $\tilde{z} = 0.2$ for (a) quadcopter ($P = 4$), (b) hexacopter ($P = 6$), and power-lift eVTOL configurations with (c) $P = 2$, (d) $P = 6$, (e) $P = 8$, and (f) $P = 12$, all evaluated at a normalized inter-propeller distance of $\tilde{L} = 1.2$.

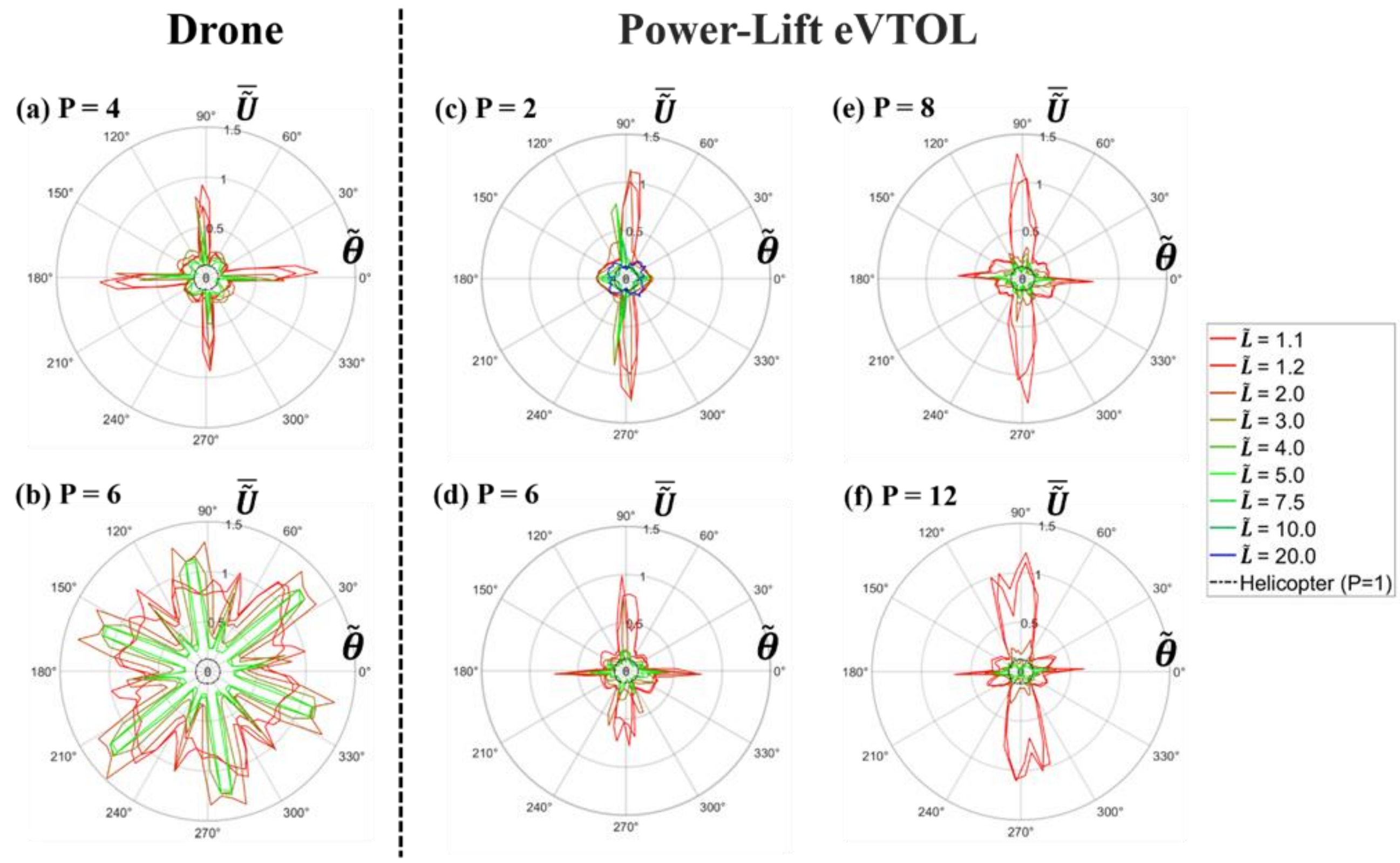

Figure 6 – Orientational distributions of mean outwash speed ($\tilde{\bar{U}}$) for (a) quadcopter ($P = 4$), (b) hexacopter ($P = 6$), and power-lift eVTOL configurations with (c) $P = 2$, (d) $P = 6$, (e) $P = 8$, and (f) $P = 12$, evaluated at a height of $\tilde{z} = 0.2$. Normalized inter-propeller distances span $\tilde{L} = 1.1$–$5$ for all cases, except for the two-propeller configuration (c), where $\tilde{L} = 1.1$–$20$. The angular reference $\tilde{\theta} = 0°$ corresponds to the $\tilde{x}$-direction.

### *3.4 Propagation range of jet-like eVTOL outwash*

To further quantify the propagation characteristics of jet-like outwash across varying propeller alignments and inter-propeller distances, Figure 7 illustrates the normalized propagation range of hazardous outwash ($\tilde{l}_{\tilde{U}>0.8}$). This parameter is defined as the diameter of the smallest circle encompassing the region where the outwash speed exceeds $\tilde{U} > 0.8$, normalized by TLOF length, ($\widetilde{RD}$), as specified in Eq. (6). For reference, results for a conventional single-rotor helicopter ($P = 1$) are also provided. The analysis reveals that decreasing the normalized inter-propeller distance ($\tilde{L}$) consistently enlarges the hazardous outwash range for all eVTOL configurations, frequently exceeding that of helicopters. Notably, among power-lift eVTOLs ($P = 2,6,8,12$), higher propeller counts at fixed $\tilde{L}$ are associated with reduced outwash propagation, with the $P = 12$ configuration at $\tilde{L} \geq 2$ exhibiting shorter hazardous ranges than the baseline helicopter.

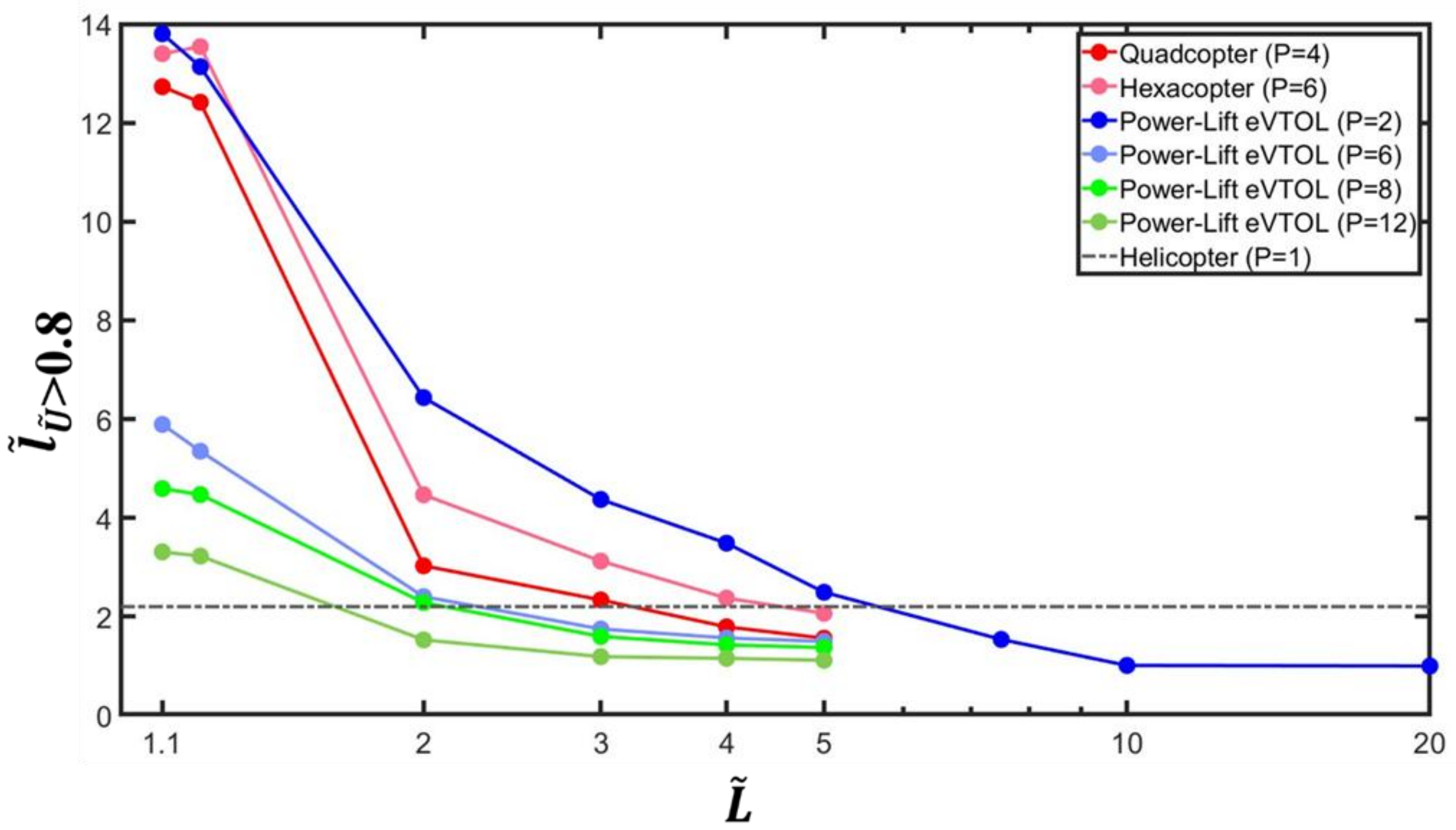

Figure 7 – Normalized propagation range of hazardous outwash ($\tilde{l}_{\tilde{U}>0.8}$), defined as the ratio between the diameter of the smallest circle enclosing the region where outwash speed exceeds $\widetilde{U} > 0.8$ and the TLOF length ($\widetilde{RD}$) specified in Eq. (6), shown as a function of normalized inter-propeller distance ($\tilde{L}$) for different eVTOL propeller configurations.

These results carry critical implications for vertiport safety and regulatory frameworks:

- First, they demonstrate that existing standards for vertiport safety areas are inadequate for eVTOL operations. For example, a power-lift eVTOL with $P = 2$ can generate hazardous outwash zones extending up to 14 times the TLOF length at $\tilde{L} = 1.1$. By contrast, the FAA EB105a standard currently specifies a safety area of only 2.5 times the control diameter (e.g., wingspan), a criterion sufficient for helicopters—whose hazardous outwash extends only about 2.2 times the TLOF length—but grossly insufficient for multi-rotor eVTOLs. This discrepancy highlights the urgent need for revising vertiport safety-area requirements to accommodate the unique outwash dynamics of distributed propulsion systems.
- Second, the findings reveal that careful geometric alignment and adequate inter-propeller spacing in power-lift eVTOLs with more than six propellers can actually limit hazardous outwash propagation to levels shorter than those of conventional helicopters. This not only underscores the complex, configuration-dependent nature of eVTOL outwash phenomena but also emphasizes that rigorous propeller alignment optimization from the design stage offers a viable pathway to fundamentally mitigate safety risks and ensure compliance with evolving regulatory guidance.

### 3.5 *Ground-level boundary layer structure of jet-like eVTOL outwash*

To gain deeper insight into the ground-level flow dynamics of jet-like outwash, this study evaluates the mean radial velocity ($\overline{\tilde{u}_r}$) as a function of height above ground ($\tilde{z}$) within the FATO region ($\widetilde{RD}/2 \leq \tilde{r} \leq \widetilde{RD}$) for various eVTOL propeller configurations and normalized inter-propeller distances ($\tilde{L}$). The results are shown in Figure 8. For quadcopters ($P = 4$), radial velocity profiles were sampled and averaged along the four inter-propeller gap orientations ($\tilde{\theta} = 0°, 90°, 180°, 270°$; Fig. 8a, corresponding to Fig. 6a). For hexacopters ($P = 6$), sampling was conducted along the six gap orientations ($\tilde{\theta} = 3\,0°, 90°, 150°, 210°, 270°, 330°$; Fig. 8b, corresponding to Fig. 6b). For power-lift eVTOLs ($P = 2,6,8,12$), analysis was focused along the longitudinal axis of the aircraft ($\tilde{\theta} = 90°$ and $270°$; Figs. 8c–f, corresponding to Figs. 6c–f).

In addition, Figure 9 quantifies two key metrics: (a) the thickness of the radial velocity boundary layer ($\tilde{h}$), defined as the height where ($\overline{\tilde{u}_r}$) decreases below 0.8, and (b) the maximum mean radial velocity $\overline{\tilde{u}}_{r_{max}}$ at the boundary layer height. Together, these results provide a systematic characterization of the vertical structure and intensity of jet-like outwash across different eVTOL architectures.

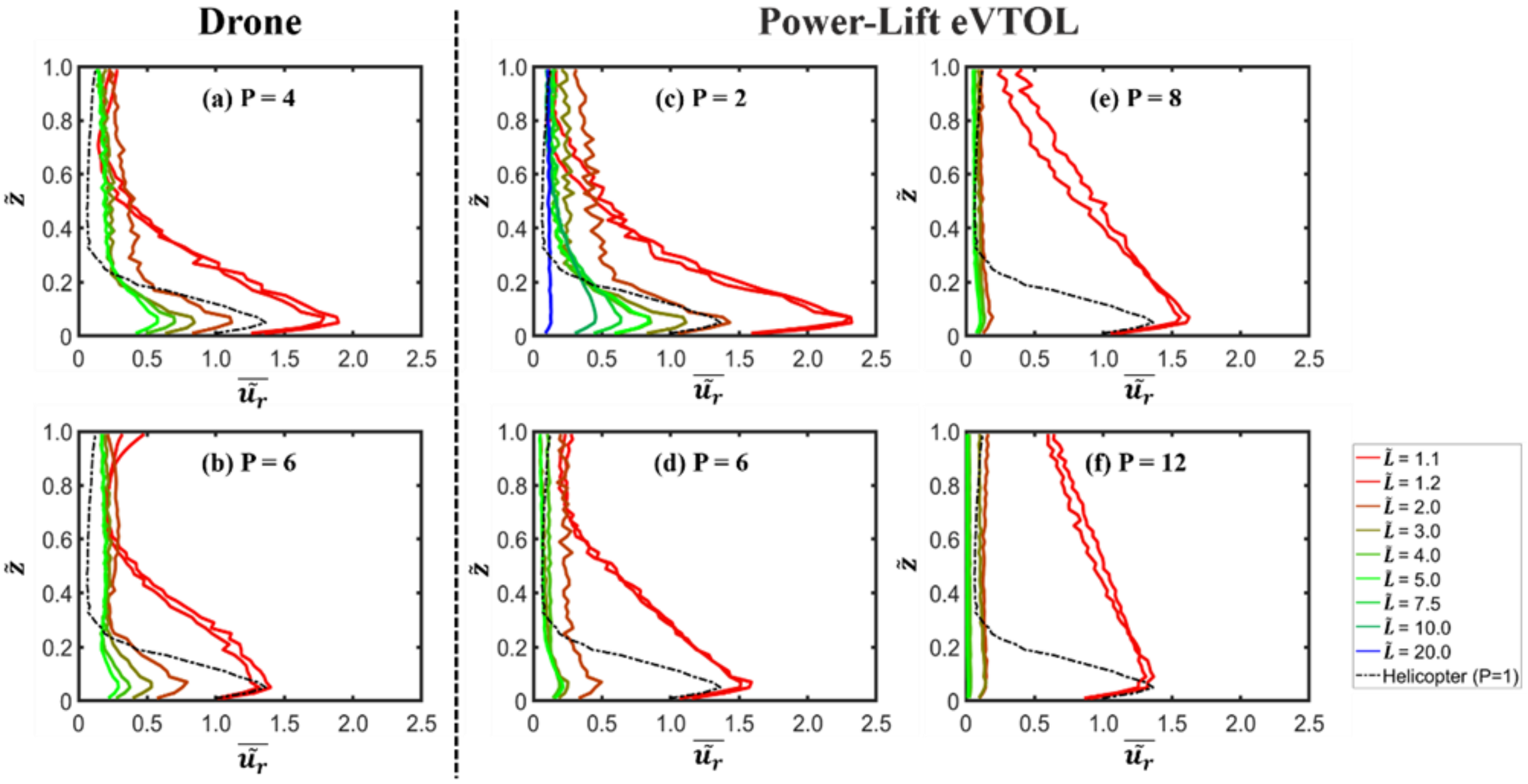

Figure 8 – Mean radial velocity ($\overline{\tilde{u}_r}$) of jet-like outwash as a function of height above ground ($\tilde{z}$) within the FATO region for (a) quadcopter ($P = 4$), (b) hexacopter ($P = 6$), (c) power-lift eVTOL ($P = 2$), (d) power-lift eVTOL ($P = 6$), (e) power-lift eVTOL ($P = 8$), and (f) power-lift eVTOL ($P = 12$). Results are shown for normalized inter-propeller distances of $\tilde{L} = 1.1$–$5$, except for the $P = 2$ configuration (c), where $\tilde{L}$ ranges from 1.1 to 20.

Figure 8 reveals that the peak outwash velocity across all eVTOL propeller configurations consistently occurs at a normalized height of $\tilde{z} \approx 0.1$. Complementary results in Figure 9 demonstrate that both the magnitude of the boundary-layer outwash speed and the thickness of the hazardous outwash zone are strongly dependent on propeller geometry and normalized inter-propeller spacing ($\tilde{L}$). Specifically, as $\tilde{L}$ increases, both the maximum mean radial velocity $\overline{\tilde{u}_{r}}_{max}$ and the hazardous boundary-layer thickness ($\tilde{h}$) decrease. For power-lift eVTOLs, an increase in propeller count reduces the near-ground maximum outwash velocity (Fig. 9b), yet simultaneously produces a thicker hazardous outwash layer (Fig. 9a). This behavior indicates that parallel jet-like outwash streams generated by high-propeller-count eVTOLs, with propellers aligned along the wingspan, tend to vertically "stack," creating a resultant flow that is slower than a single concentrated jet yet characterized by greater vertical extent. Remarkably, this vertical stacking effect produces near-ground outwash velocities comparable to those of helicopters at low $\tilde{L}$, but with significantly enhanced thickness. To the authors' knowledge, such collective vertical stacking of outwash flows in multi-rotor eVTOLs has not been reported in previous literature.

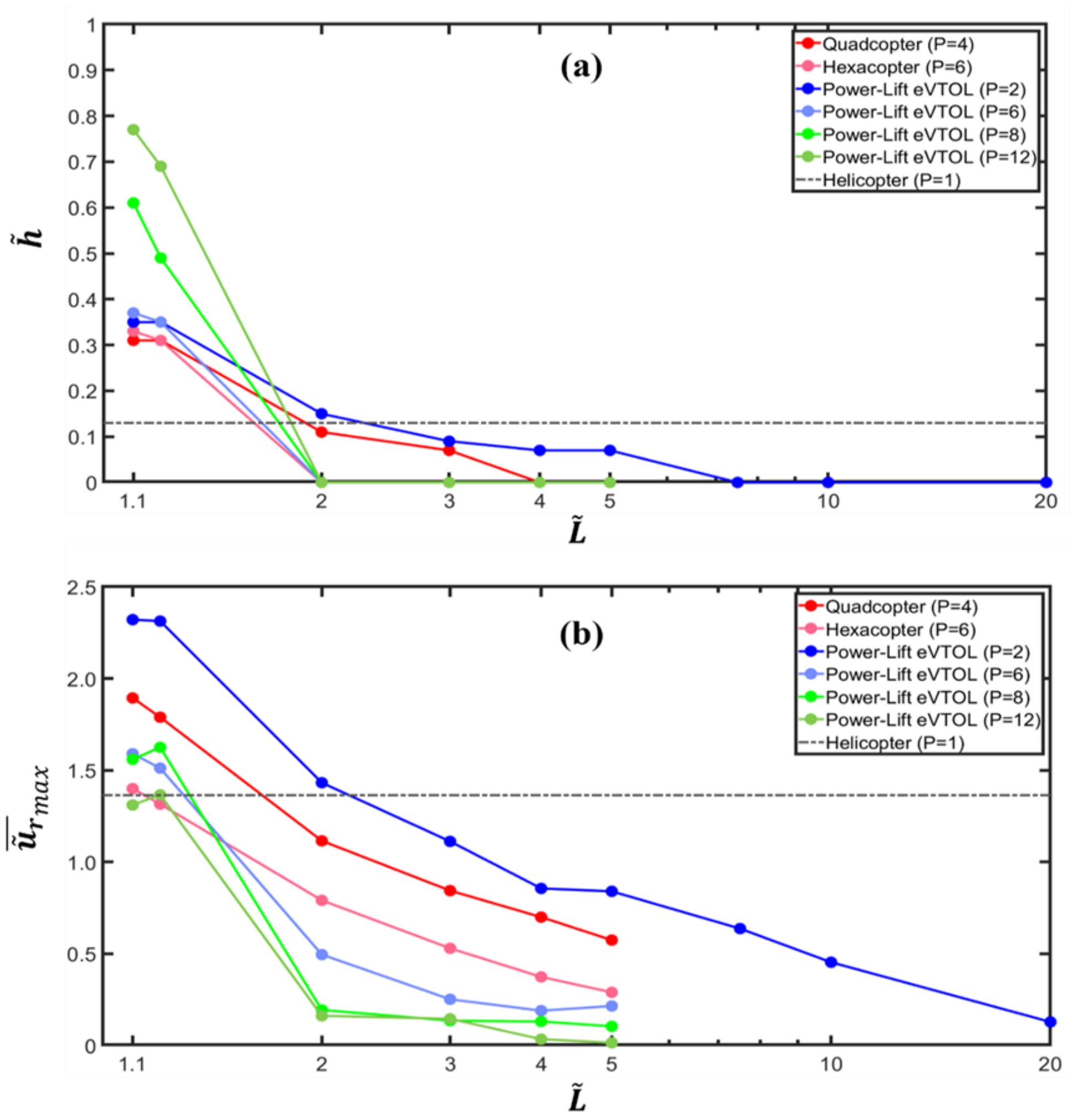

Figure 9 – (a) Thickness of the radial velocity boundary layer ($\tilde{h}$) and (b) maximum mean radial velocity ($\overline{\tilde{u}_{r}}_{max}$) within the boundary layer, shown as functions of normalized inter-propeller distance ($\tilde{L}$) for different eVTOL propeller configurations.

These findings yield critical design implications: effective outwash mitigation must be tailored to the specific rotor configuration of the eVTOL. For high-propeller-count power-lift designs, the ground-level outwash velocity is comparatively weaker—similar to that of helicopters—but the hazardous outwash layer extends higher, necessitating taller barriers to provide adequate protection. Conversely, low-propeller-count power-lift eVTOLs generate more intense ground-level outwash, yet the hazardous region remains relatively shallow, allowing shorter barriers to achieve effective mitigation. These insights are essential for developing cost-efficient and scalable outwash control strategies at vertiports, and they provide actionable guidance for optimizing propeller alignment in future eVTOL design practices.

### *3.6 Developing outwash mitigation strategies based on outwash flow patterns*

To rigorously assess whether insights from eVTOL outwash studies—particularly those concerning jet-like outwash intensity, orientation, propagation range, and near-ground boundary layer structure—can effectively inform the design of cost-efficient vertiport mitigation devices, this analysis examines a representative power-lift eVTOL with $P = 2$ and $\tilde{L} = 2$. In the absence of mitigation, the hazardous outwash region ($\widetilde{U} > 0.8$) extends to nearly 6.45 times the TLOF length (Fig. 7), with the strongest jet-like flows aligned along the aircraft's longitudinal axis (Fig. 6c) and a boundary layer thickness of 0.15 (Fig. 9a). To counteract this hazard, two blast deflectors were installed at the FATO boundary, oriented perpendicular to the dominant outwash direction. Three deflector designs were evaluated: (i) a flat plate, (ii) an inclined plate at 45°, and (iii) a hybrid structure with the lower three-quarters inclined and the upper quarter flat. Each configuration was tested at multiple heights ($\widetilde{H} = 0.2, 0.15, 0.1, 0.05$). Figure 10 illustrates the impact of all three deflector types at $\widetilde{H} = 0.2$, while Figure 11 details the performance of inclined deflectors across all tested heights. Table 1 summarizes the normalized propagation range of hazardous outwash ($\tilde{l}_{\widetilde{U}>0.8}$), as defined in Fig. 7, distinguishing between near-ground regions ($\tilde{z} \leq 0.5$) and the full airspace ($\tilde{z} \in \mathbb{R}^+$).

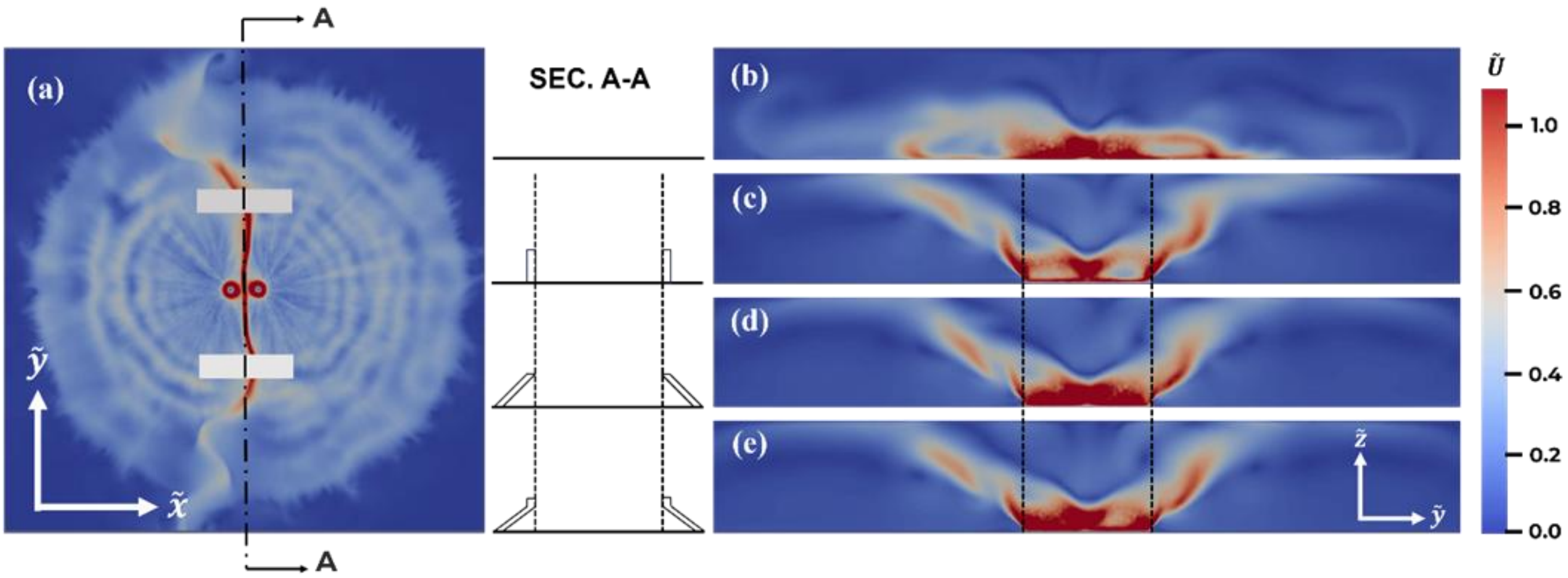

Figure 10 – Outwash flow speed distribution: (a) baseline case without deflector at $\tilde{z} = 0.2$; (b) cross-sectional profile along $\tilde{x} = 0$ from Fig. 10a; (c) with flat deflector installed at $\widetilde{H} = 0.2$ ($\tilde{x} = 0$); (d) with inclined deflector installed at $\widetilde{H} = 0.2$ ($\tilde{x} = 0$); and (e) with hybrid deflector (lower 3/4 inclined, upper 1/4 flat) installed at $\widetilde{H} = 0.2$ ($\tilde{x} = 0$).

The results demonstrate that for deflectors with heights of $\widetilde{H} = 0.2$ and $\widetilde{H} = 0.15$, the hazardous outwash near the ground ($\tilde{z} \leq 0.5$) is consistently confined within 2.5 times the TLOF length (Figs. 10c–e, Figs. 11a–b, Table 1), fully satisfying the FAA EB105a safety requirements. By contrast, shorter deflectors ($\widetilde{H} = 0.1, 0.05$) fail to contain ground-level outwash within these limits, emphasizing that the deflector height must at least match or exceed the pre-mitigation outwash thickness (0.15) to ensure effective hazard reduction. In all cases, the installed deflectors redirect the hazardous outwash upward into the atmosphere and laterally away from the eVTOL body. The mitigation effectiveness is governed primarily by the geometry and height of the deflector, as these parameters directly determine the deflection angle and, consequently, the overall propagation range of hazardous outwash in the airspace. For example, at $\widetilde{H} = 0.2$, the inclined deflector (Fig. 10d) proves most effective, limiting the propagation range to $\tilde{l}_{\tilde{U}>0.8}\big|_{\tilde{z}\in\mathbb{R}^+} = 2.73$, compared to 3.13 for the flat variant (Fig. 10c, Table 1). Reducing the inclined deflector height elongates the propagation range, reaching $\tilde{l}_{\tilde{U}>0.8}\big|_{\tilde{z}\in\mathbb{R}^+} = 4.07$ at $\widetilde{H} = 0.05$ (Fig. 11d). At this minimal height, the deflector no longer provides meaningful upward redirection, and the ground-level outwash propagation becomes indistinguishable from the total airspace result, confirming that insufficient height renders the device ineffective (Table 1).

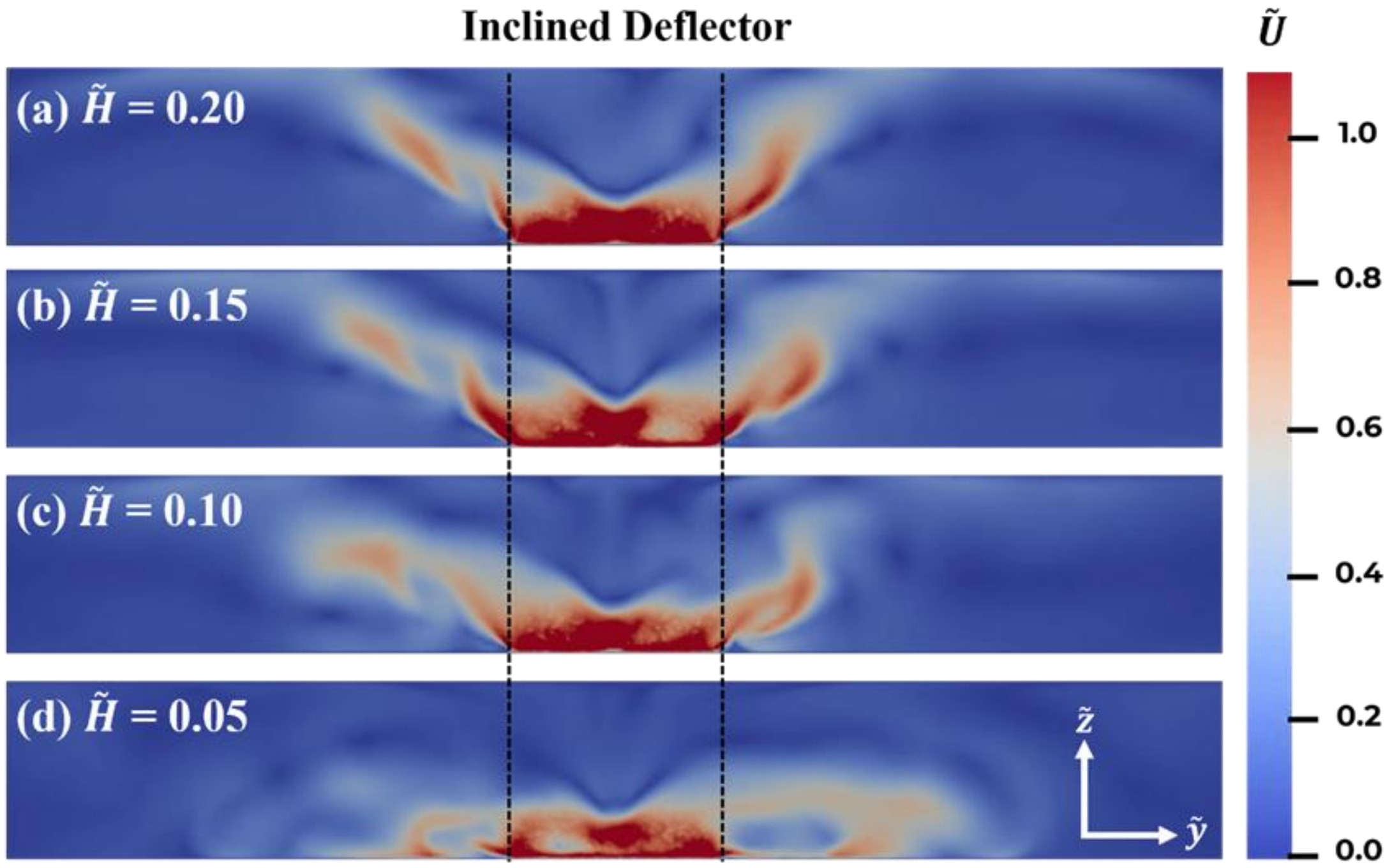

Figure 11 – Outwash flow speed distribution at the plane of $\tilde{x} = 0$ following installation of inclined deflectors with varying heights: (a) $\tilde{H}$ = 0.2; (b) $\tilde{H}$ = 0.15; (c) $\tilde{H}$ = 0.1; and (d) $\tilde{H}$ = 0.05. The comparison highlights how deflector height influences the containment and upward redirection of hazardous outwash.

Assessing vertiport area requirements, when the necessary buffer zone is defined as a circle of diameter $\tilde{l}_{\tilde{U}>0.8}\big|_{\tilde{z}\in\mathbb{R}^+}$, the installation of mitigation devices leads to dramatic spatial savings. The optimal inclined deflector ($\widetilde{H} = 0.2$) achieves up to an 82% reduction in required area, while even the least effective case—the flat deflector at $\widetilde{H} = 0.05$—still provides a 58% reduction. These results clearly demonstrate that carefully engineered outwash mitigation devices, designed with precise knowledge of eVTOL outwash dynamics, can substantially reduce vertiport footprint with minimal structural additions. For example, for eVTOLs equipped with propellers of 5 m diameter, a blast deflector as small as 25 cm in height ($\widetilde{H} = 0.05$) is sufficient to meaningfully constrain the outwash buffer zone. Such compact and cost-efficient solutions not only lower construction expenses but also enable scalable deployment of vertiports in dense urban environments.

Table 1. Normalized propagation range of hazardous outwash ($\tilde{l}_{\tilde{U}>0.8}$), distinguishing near-ground regions ($\tilde{z} \leq 0.5$) from the entire airspace ($\tilde{z} \in \mathbb{R}^+$)

| **Normalized Hazardous Outwash Propagation Range ($\tilde{l}_{\tilde{U}>0.8}$; TLOF unit)** | | | |
|---|---|---|---|
| | **Flat Deflector** | **Inclined Deflector** | **Hybrid Deflector** |
| **$\tilde{H}$ = 0.20** | **2.13 / 3.13** | **2.33 / 2.73** | **2.30 / 2.83** |
| **$\tilde{H}$ = 0.15** | **2.25 / 3.20** | **2.37 / 3.00** | **2.45 / 3.87** |
| **$\tilde{H}$ = 0.10** | **2.60 / 3.77** | **2.60 / 3.07** | **2.63 / 3.67** |
| **$\tilde{H}$ = 0.05** | **4.17 / 4.17** | **4.07 / 4.07** | **3.20 / 3.20** |
| **No Deflector** | **6.45 / 6.45** | | |
| **Data are expressed in the form of ($\tilde{l}_{\tilde{U}>0.8}\|_{\tilde{z}\leq 0.5}$ / $\tilde{l}_{\tilde{U}>0.8}\|_{\tilde{z}\in\mathbb{R}^+}$)** | | | |

This study identifies several promising directions for future research. A central question concerns the determination of the minimum effective deflector height across different geometrical designs—a challenge underscored by the findings in Table 1, where the normalized hazardous outwash propagation range ($\tilde{l}_{\tilde{U}>0.8}|_{\tilde{z}\in\mathbb{R}^+}$) for the hybrid deflector does not follow a monotonic trend with decreasing deflector height ($\widetilde{H}$). This nonlinearity highlights the complex aerodynamic interactions between jet-like outwash and blast deflector geometries. Further investigation is needed to identify which configurations achieve the greatest reduction in hazardous outwash with the least structural height, as well as to develop strategies that not only attenuate outwash speed but also redirect its trajectory so that all hazardous flows remain confined within FAA EB105a safety limits. Collectively, these avenues represent critical opportunities for advancing vertiport safety and for refining outwash mitigation strategies into more compact, efficient, and scalable engineering solutions.

## 4. Conclusions and outlook

This study conducted a comprehensive numerical investigation to characterize the outwash phenomena generated by multi-rotor eVTOL aircraft in hover, with a particular focus on how propeller geometry and alignment influence hazardous flow formation. Employing nondimensional RANS simulations with the k–ω SST turbulence model, the analysis distinguished eVTOL jet-like outwash from conventional helicopter wakes, revealing unique boundary-layer structures, directional intensification, and propagation patterns. A systematic evaluation of propeller count and spacing further clarified how these parameters shape outwash behavior, while the effectiveness of blast deflectors as mitigation devices was assessed in terms of reducing vertiport buffer zone requirements.

The principal findings can be summarized as follows:

(a) Inherent jet-like outwash in multi-rotor eVTOLs: Jet-like outwash is an intrinsic characteristic of multi-rotor eVTOLs under standard inter-propeller spacing. It originates from the pressure maximum forming between adjacent rotors, which drives an amplified pressure gradient perpendicular to the alignment axis, intensifying outwash in the inter-propeller direction.
(b) Effect of spacing and rotor count on hazard severity: Power-lift eVTOLs with small inter-propeller distances and fewer propellers generate intense, highly directional outwash jets, with propagation ranges far exceeding the safety envelopes defined by FAA EB105a standards. These results underscore the urgent need for type-specific safety guidelines and revised vertiport planning criteria.
(c) Collective outwash dynamics in high-propeller-count designs: For power-lift eVTOLs with many propellers aligned along the wingspan, the outwash streams merge into a collective longitudinal jet characterized by reduced peak velocity but a vertically thickened boundary layer. While the shortened propagation range reduces far-field hazards, the thicker near-ground hazard layer demands tailored mitigation strategies beyond current helicopter-based safety standards.
(d) Engineering-based mitigation and infrastructure optimization: Modular blast deflectors, strategically designed based on predicted outwash intensity, orientation, propagation range, and boundary-layer thickness, can shrink required vertiport safety areas by up to 82% with minimal structural intervention. Such solutions enable practical, scalable, and cost-effective infrastructure while ensuring compliance with safety regulations.

By integrating aerodynamic insights, regulatory frameworks, and engineering interventions, this study provides a framework for aligning eVTOL aircraft design with vertiport infrastructure and urban safety requirements. The findings highlight promising pathways for future research, including optimization of deflector geometries and the development of robust, simulation-informed outwash mitigation strategies. Collectively, this work advances the aerodynamic foundations of urban air mobility and supports the safe, efficient, and sustainable deployment of eVTOL systems.

## Acknowledgments

This work was partially supported by the Headquarters of University Advancement at National Cheng Kung University, funded by the Ministry of Education, Taiwan, and by the National Science and Technology Council (NSTC), Taiwan, under grant numbers 108-2218-E-006-028-MY3, 111-2221-E-006-102-MY3, and 114-2221-E-006-071. The authors used Perplexity AI to assist with improving the readability, grammar, and language of the manuscript. All technical content, analyses, interpretations, references, and conclusions were developed, reviewed, and verified by the authors, who assume full responsibility for the accuracy and integrity of the paper.

## Conflicts of interest

Yen-Chen Chen is the founder and Chief Executive Officer of Peniel Infrastructure LTD., which has a commercial interest in outwash-mitigation technologies related to this study.

## CRediT authorship contribution statement

**Yen-Chen Chen**: Writing – review & editing, Writing – original draft, Project administration, Methodology, Investigation, Formal analysis, Conceptualization, Visualization, Validation, Software, Data curation. **Chih-Che Chueh**: Writing – review & editing, Conceptualization, Supervision, Resources.

## Data availability

The data that support the findings of this study are available from the corresponding author upon reasonable request.

## Nomenclature

| | |
|---|---|
| $A$ | cross-sectional area of propeller outlet |
| $D$ | propeller diameter, characteristic length scale for nondimensionalization |
| $D_a$ | propeller diameter $a$ in comparative sizing formula |
| $D_b$ | propeller diameter $b$ in comparative sizing formula |
| $F_1$ | blending function in SST model |
| $\tilde{h}$ | nondimensional thickness of radial velocity boundary layer |
| $\tilde{H}$ | nondimensional blast deflector height |
| $\tilde{k}$ | dimensionless turbulence kinetic energy, $k/U^2$ |
| $\tilde{\omega}$ | nondimensional specific turbulence dissipation rate, $\omega D/U$ |
| $L$ | inter-propeller center distance |
| $\tilde{L}$ | normalized inter-propeller distance, $L/D$ |
| $W$ | lateral propeller spacing |
| $\tilde{W}$ | normalized lateral propeller spacing, $W/D$ |
| $\tilde{l}_{\tilde{U}>0.8}$ | normalized propagation range of hazardous outwash |
| $P$ | number of propellers |
| $P_a$ | number of propellers $a$ in comparative sizing formula |
| $P_b$ | number of propellers $b$ in comparative sizing formula |
| $\tilde{P}_k$ | nondimensional production term of turbulent kinetic energy |
| $\tilde{p}$ | nondimensional pressure, $p/(\rho U^2)$ |
| $\nabla\tilde{p}$ | nondimensional pressure gradient |
| $\rho$ | fluid density |
| $\mu$ | dynamic viscosity |
| $\tilde{\nu}$ | nondimensional kinematic viscosity, $\nu/(UD)$ |
| $\tilde{\nu}_t$ | nondimensional turbulent (eddy) viscosity, $\nu_t/(UD)$ |
| $Re_D$ | Reynolds number, $\rho UD/\mu$ |
| $\mathbb{R}^+$ | set of all positive real numbers, vertical coordinate domain |
| $RD$ | rotor diameter; diameter of smallest circle encompassing all rotors |
| $\widetilde{RD}$ | nondimensional rotor diameter, $RD/D$ |
| $T$ | thrust force of a single propeller |
| $T/A$ | disk loading of the propeller |
| $U$ | reference flow velocity at the rotor disk, used to normalize velocity fields |
| $\tilde{U}$ | nondimensional speed (normalized by disk velocity) |
| $\overline{\tilde{U}}$ | mean nondimensional outwash speed |
| $\tilde{u}_i$ | nondimensional velocity component, $u_i/U$ |
| $\tilde{u}_i'$ | fluctuating component of $\tilde{u}_i$ |
| $\overline{\widetilde{u_r}}$ | mean normalized radial velocity of outwash |
| $\overline{\tilde{u}}_{r_{max}}$ | maximum mean normalized radial velocity at boundary layer |
| $\tilde{t}$ | Nondimensional time, $tU/D$ |
| $\tilde{x}_i$ | nondimensional spatial coordinate, $x_i/D$ |
| $\tilde{x}$ | nondimensional wingspan direction of power-lift eVTOL |
| $\tilde{y}$ | nondimensional longitudinal direction of power-lift eVTOL |
| $\tilde{z}$ | nondimensional vertical coordinate (height) |
| $\tilde{r}$ | nondimensional radial coordinate |
| $\tilde{\theta}$ | nondimensional azimuthal coordinate |
| $\alpha$ | empirical constant in the k–ω SST model equations |
| $\beta$ | empirical constant in the k–ω SST model equations |
| $\tilde{\beta}$ | empirical constant in the k–ω SST model equations |
| $\tilde{\sigma}_k$ | empirical diffusion coefficient for k |
| $\tilde{\sigma}_\omega$ | empirical diffusion coefficient for ω |
| $\tilde{\sigma}_{\omega 2}$ | empirical diffusion coefficient for ω |
| $\Omega$ | rotational speed of propeller |
| $\Omega_a$ | rotational speed $a$ in comparative sizing formula |
| $\Omega_b$ | rotational speed $b$ in comparative sizing formula |

*Abbreviations*

| | |
|---|---|
| $AAM$ | advanced air mobility |
| $CAA$ | Civil Aviation Authority (UK) |
| $CFD$ | computational fluid dynamics |
| $EB105a$ | Engineering Brief No. 105A (vertiport safety guidelines) |
| $eVTOL$ | electric vertical takeoff and landing aircraft |
| $FATO$ | final approach and takeoff area |
| $FAA$ | Federal Aviation Administration |
| $NACA$ | National Advisory Committee for Aeronautics |
| $NASA$ | National Aeronautics and Space Administration |

$RANS$ Reynolds-Averaged Navier–Stokes Equations

$SST$ Shear Stress Transport (turbulence model)

$TLOF$ touchdown and liftoff area

$UAM$ urban air mobility

$UAV$ unmanned aerial vehicle

$VTOL$ vertical takeoff and landing